\documentclass[aps,groupedaddress, superscriptaddress,showpacs,amsmath,amssymb,twocolumn,pra]{revtex4-1}
\newcounter{defcounter}
\usepackage{bm}
\usepackage[dvips]{graphicx}
\usepackage{wrapfig,subfigure}
\usepackage{amssymb}
\usepackage{color}
\usepackage[normalem]{ulem}
\usepackage{amsmath}
\usepackage{epstopdf}
\newcommand{\bn}{\boldsymbol{n}}
\newcommand{\Kb}{{\boldsymbol{K}}}
\newcommand{\nb}{{\boldsymbol{n}}}
\newcommand{\rb}{\boldsymbol{r}}

\newcommand{\vb}{{\boldsymbol{v}}}

\newcommand{\bsigma}{\boldsymbol{\sigma}}

\newcommand{\be}{\begin{equation}}
\newcommand{\ee}{\end{equation}}

\begin{document}
\title{Multi-time correlators in continuous measurement of qubit observables}
\author{Juan Atalaya}
\affiliation{Department of Electrical and Computer Engineering, University of California, Riverside, CA 92521, USA}
\author{Shay Hacohen-Gourgy}
\affiliation{Quantum Nanoelectronics Laboratory, Department of Physics, University of California, Berkeley CA 94720, USA}
\affiliation{Center for Quantum Coherent Science, University of California, Berkeley CA 94720, USA.}
\author{Leigh S. Martin}
\affiliation{Quantum Nanoelectronics Laboratory, Department of Physics, University of California, Berkeley CA 94720, USA}
\affiliation{Center for Quantum Coherent Science, University of California, Berkeley CA 94720, USA.}
\author{Irfan Siddiqi}
\affiliation{Quantum Nanoelectronics Laboratory, Department of Physics, University of California, Berkeley CA 94720, USA}
\affiliation{Center for Quantum Coherent Science, University of California, Berkeley CA 94720, USA.}
\author{Alexander N. Korotkov}
\affiliation{Department of Electrical and Computer Engineering, University of California, Riverside, CA 92521, USA}
\date{\today}
\begin{abstract}
We consider multi-time correlators for output signals from linear detectors, continuously measuring several qubit observables at the same time. Using the quantum Bayesian formalism, we show that for unital (symmetric) evolution in the absence of phase backaction, an $N$-time correlator can be expressed as a product of two-time correlators when $N$ is even. For odd $N$, there is a similar factorization, which also includes a single-time average. Theoretical predictions agree well with experimental results for two detectors, which simultaneously measure non-commuting qubit observables.
\end{abstract}
\maketitle

{\it Introduction.}---Partial and continuous quantum measurements (CQMs) have recently attracted a significant attention within the quantum information community. They have been discussed theoretically for a long time~\cite{Kraus1983, Caves1986, Menskii1998, Belavkin1992, Braginsky-book, Aharonov1988, Diosi1988, Dalibard1992, Carmichael1993, Wiseman1993, Korotkov1999}, and renewed interest has been motivated by the rapid progress with superconducting qubits, which are currently the main experimental system for realization of CQMs \cite{Katz2006, Laloy2010, Vijay2012, Hatridge2013, Murch2013, deLange2014, Campagne2014}. The main contribution from the theory and experiments on CQMs to fundamental aspects of quantum physics is a clear understanding of quantum evolution in the process of the state collapse due to measurement. Besides that, we now understand quantum dynamics in some peculiar processes like uncollapse \cite{Korotkov2006, Katz2008} and simultaneous measurement of non-commuting observables \cite{Ruskov2010, Shay2016}. There is also a growing interest in the use of CQMs for quantum computing applications, such as quantum feedback \cite{Wiseman1993fb, Ruskov2002, Sayrin2011, Vijay2012, deLange2014, Patti2017}, rapid state purification \cite{Jacobs2003}, entanglement by continuous measurement \cite{Ruskov2003, Riste2013, Roch2014} and quantum error correction \cite{Ahn2002, Atalaya2017}.

A recent experiment \cite{Shay2016} opened a way to experimental verification of theoretical predictions related to simultaneous  measurement of non-commuting observables. In particular, two-time correlators for the output signals from two detectors measuring two observables of a qubit, have been calculated in Ref.\ \cite{Atalaya2017corr}, and the theoretical results showed a good agreement with experimental data. Similarly, the theoretical results for two-time state correlators have been compared with experiment in Ref.\ \cite{Chantasri2017}.

In this work we again consider temporal correlations for signals from linear detectors, simultaneously measuring  non-commuting observables of a qubit. However, now the number of detectors and observables is arbitrary, observables can change in time, and we extend the previous analysis of two-time correlators \cite{Korotkov2001sp, Jordan2005,  Atalaya2017corr} to multi-time correlators, thus fully  describing statistics of the output signals. Our analysis also extends the recent result \cite{Diosi2016} for  correlations between sequential infinitesimally weak measurements in the absence of evolution.

Let us consider a qubit, which is weakly coupled to $N_{\rm d}$ linear detectors, measuring simultaneously and continuously an arbitrary set of qubit observables $\sigma_{\ell} \equiv \nb_\ell \bsigma$, where $\ell=1,2,...\, N_{\rm d}$, the unit vector $\nb_\ell =(n_{x,\ell}, n_{y,\ell}, n_{z,\ell})$ is the $\ell$th measurement axis direction on the Bloch sphere, and $\bsigma = (\sigma_x,\sigma_y,\sigma_z)$ is the vector of Pauli matrices. For the output signals $I_{\ell}(t)$ from the detectors, the multi-time correlators of interest are
\begin{align}
\label{eq:Kn-def}
K_{\ell_1 \ell_2...\ell_N}(t_1,t_2,...t_N)\equiv \big\langle I_{\ell_N}(t_N)\cdots I_{\ell_2}(t_2)I_{\ell_1}(t_1)\big\rangle,
\end{align}
where averaging is over the ensemble of realizations, we assume that the time  arguments are ordered as $t_1 < t_2 < ... < t_N$, and $N$ can be smaller, equal, or larger than $N_{\rm d}$. As we show in this paper, for unital evolution (with symmetry between qubit states $|0\rangle$ and $|1\rangle$) in the absence of phase backaction from measurement, the $N$-time correlator (\ref{eq:Kn-def}) has a quite simple form. Rather surprisingly, for even $N$ it factorizes into a product of $N/2$ sequential two-time correlators, so that the qubit evolution between $t_{N-2k}$ and $t_{N-2k+1}$ does not affect the correlator (\ref{eq:Kn-def}). For odd $N$, there is a similar product, which also includes the average signal at the earliest time, $\langle I_{\ell_1}(t_1) \rangle$. In this paper we also compare the theoretical predictions with experimental data for a two-detector
configuration similar to Ref.\ \cite{Shay2016}.

Our results are useful for parameter estimation via correlators (see \cite{Atalaya2017corr}) and noise characterization as a tool for diagnosing sources of fluctuations in multiqubit systems, with multi-observable correlators probing the dynamics within the whole Hilbert space. Our results are also useful for analysis of error syndromes in quantum error correction codes based on continuous measurements. In particular, the theory presented here has been implicitly used (without any discussion or formulas) in Ref.\ \cite{Atalaya2017} for  error analysis in the four-qubit Bacon-Shor code operated with continuous measurements; the parity operators for that code correspond to non-commuting observables of the gauge qubit.

{\it The quantum Bayesian formalism.}---A simultaneous continuous measurement of the qubit observables $\sigma_{\ell}$ produces the normalized output signals~\cite{Korotkov1999, Korotkov2001-16,Ruskov2010}
\begin{align}
\label{eq:I-ell}
I_\ell(t) &= {\rm Tr}[{\sigma_\ell}\rho(t)] + \sqrt{\tau_\ell}\, \xi_\ell(t)
             = \bn_\ell \rb(t) + \sqrt{\tau_\ell}\, \xi_\ell(t),
\end{align}
where the Bloch vector $\rb=(x,y,z)$ for the qubit density matrix $\rho$ is defined via parametrization  $\rho = (\openone + x\sigma_x + y\sigma_y + z\sigma_z)/2$ and $\tau_\ell$ denotes the measurement  (collapse) time needed for informational signal-to-noise ratio of 1 for the $\ell$th-measurement channel. We use the Markovian quantum Bayesian theory (assuming the ``bad cavity limit''), so the noises $\xi_\ell(t)$ are assumed to be white, Gaussian and uncorrelated, with two-time correlators
\begin{align}
\label{eq:noise-corr}
\langle \xi_\ell(t)\, \xi_{\ell'}(t')\rangle = \delta_{\ell\ell'}\delta(t-t').
\end{align}
We also assume that the quantum backaction due to measurement is only of the informational type, so that there is no phase backaction \cite{Korotkov2001-16}; in a circuit QED experiment this requires that the optimal microwave quadrature is amplified by a phase-sensitive amplifier. Then the quantum Bayesian equation for the qubit state evolution (in It\^o interpretation)  is ~\cite{Korotkov2001-16,Ruskov2010,Wiseman1993}
\begin{align}
\label{eq:Bayesian-eq}
\dot\rb = \Lambda_{\rm ens}(\rb - \rb_{\rm st}) + \sum_{\ell=1}^{N_{\rm d}} \frac{\nb_\ell - (\nb_\ell \rb)\, \rb}{\sqrt{\tau_\ell}}\, \xi_\ell(t),
\end{align}
where the second term depends on the noisy outputs $I_\ell (t)$, while the  first term describes ensemble-averaged evolution, characterized by a 3$\times$3  matrix $\Lambda_{\rm ens}$ and quasistationary state  $\rb_{\rm st}$ ($\Lambda_{\rm ens}$, $\rb_{\rm st}$, $\nb_\ell$, and $\tau_\ell$ can all be time-dependent). Note that
\begin{align}
\label{eq:ens-eq}
      \dot{\rb}_{\rm ens} = \Lambda_{\rm ens}(\rb_{\rm ens} - \rb_{\rm st}),
\end{align}
is the most general form of a linear Markovian evolution of the ensemble-averaged qubit state $\rb_{\rm ens}=\langle \rb \rangle$, which directly corresponds to the Lindblad-form equation \cite{GardinerBook} $\dot\rho_{\rm ens} = -(i/\hbar) [H_{\rm q},\rho_{\rm ens}] + \mathcal{L}[\rho_{\rm ens}]$, where qubit Hamiltonian $H_{\rm q}$ describes Rabi oscillations, while $\mathcal{L}$ accounts for environmental decoherence and measurement-induced ensemble dephasing. In particular, the measurement contributes $\mathcal{L}_{\rm m}[\rho]=\sum_\ell \Gamma_\ell\left[\sigma_\ell\rho\sigma_\ell - \rho\right]/2$, where $\Gamma_\ell=1/2\eta_\ell\tau_\ell$ and $\eta_\ell$ is the quantum efficiency of $\ell$th detector \cite{Korotkov2001-16,Ruskov2010}.

An important special case is unital evolution, for which $\rb_{\rm st}=0$, so that a fully mixed qubit state, $\rb_{\rm ens}=0$, does not evolve. Unital evolution essentially means the symmetry between the qubit states $|0\rangle$ and $|1\rangle$. In experiments with superconducting qubits this symmetry is usually broken by energy relaxation; however, there is an approximate symmetry if Rabi oscillations are much faster than energy relaxation. In particular, in the experiment \cite{Shay2016} on simultaneous measurement of non-commuting observables and for the data used in this paper, the evolution of the effective qubit is practically unital \cite{Atalaya2017corr}. Let us denote the solution of Eq.~\eqref{eq:ens-eq} as $\rb_{\rm ens}(t|\rb_{\rm 0},t_{\rm 0})$, where $\rb_{\rm 0}$ is an initial condition at time $t_{\rm 0}$. For unital evolution
\begin{align}
\label{eq:unital-prop}
\rb_{\rm ens}\big(t\big|-\rb_{\rm 0},t_{\rm 0}\big) = -\rb_{\rm ens}\big(t\big|\rb_{\rm 0},t_{\rm 0}\big).
\end{align}

If phase backaction is included, then we need to add into Eq.\ (\ref{eq:Bayesian-eq})  the term $\sum_{\ell} \tau_\ell^{-1/2} {\cal K}_\ell (\nb_\ell \times \rb) \, \xi_\ell (t)$, where the coefficient ${\cal K}_\ell$ parametrizes the relative strength of phase backaction \cite{Korotkov2001-16} (${\cal K}_\ell=\tan \phi_\ell$, where $\phi_\ell$ is the angle between the amplified and optimal quadratures). The ensemble dephasing rate is then $\Gamma_\ell=(1+{\cal K}_\ell^2)/2\eta_\ell \tau_\ell$. However, as mentioned above, in this paper we assume ${\cal K}_\ell =0$.

{\it Collapse recipe.}---As shown in Section A of Supplemental Material \cite{Supplement}, in the absence of phase backaction (${\cal K}_\ell =0$, unitality is not needed), the qubit evolution (\ref{eq:Bayesian-eq}) leads to the following simple recipe for calculation of the multi-time correlators (\ref{eq:Kn-def}). The correct result can be obtained by replacing   actual continuous measurement with projective measurement of operators $\sigma_{\ell_k}$ at time moments $t_k$ ($k=1, 2, ...\, N$), while the qubit evolution at $t\neq t_k$ is replaced with the ensemble-averaged evolution. This ``collapse recipe'' was proven in Ref.\ \cite{Korotkov2001sp} for $N_{\rm d}=1$ and $N=2$, and also in Ref.\ \cite{Atalaya2017corr} for $N=N_{\rm d}=2$. In Section A of Supplemental Material we prove it for arbitrary $N$ and $N_{\rm d}$.

Using this recipe, in Eq.\ (\ref{eq:Kn-def}) we have $2^N$ combinations of discrete outcomes, $I_{\ell_k}=\pm 1$, each of them correspondingly collapsing the qubit state to the point $\pm\nb_{\ell_k}$ on the Bloch sphere (an eigenstate of $\sigma_{\ell_k}$). Each combination contributes the value $\prod_{k=1}^N I_{\ell_k}$ to the $N$-time correlator, with the weight equal to the probability of such combination of outcomes. In this way we obtain
\begin{align}
\label{eq:collapse-recipe}
& \hspace{0.0cm} K_{\ell_1...\ell_N}(t_1,...\, t_N) = \sum\nolimits_{\{ I_{\ell}=\pm 1\}}^{2^N} p(I_{\ell_N} \big|I_{\ell_{N-1}})
    \nonumber \\
& \hspace{0.6cm} \times p(I_{\ell_{N-1}} \big| I_{\ell_{N-2}})    \cdots
  p(I_{\ell_2} \big| I_{\ell_1}) \, p(I_{\ell_1}) \prod\nolimits_{k=1}^N I_{\ell_k},
\end{align}
where
$p(I_{\ell_k} \big| I_{\ell_{k-1}})\equiv p(I_{\ell_k},t_k \big|I_{\ell_{k-1}},t_{k-1})$ is the probability to obtain projective result $I_{\ell_k}$ at time $t_k$ if at time $t_{k-1}$ the  result was $I_{\ell_{k-1}}$  (for brevity we omit time moments in the notation), while $p(I_{\ell_1})\equiv p(I_{\ell_1},t_1)$ is the probability to obtain projective result $I_{\ell_1}=\pm 1$ at time $t_1$. It is easy to see that
\begin{align}
\label{eq:P-xi}
p (I_{\ell_1}) = \frac{1 + I_{\ell_1}{\rm Tr}[\sigma_{\ell_1}\rho_{\rm ens}(t_1)]}{2}= \frac{ 1 + I_{\ell_1} \nb_{\ell_1}\rb_{\rm ens}(t_1)}{2},
\end{align}
where $\rb_{\rm ens}(t_1)$ is the qubit state at the time $t_1$. If the qubit is prepared in a state $\rb_{\rm in}$ at the time $t_{\rm in}<t_1$, then $\rb_{\rm ens}(t_1)$ is obtained via Eq.~\eqref{eq:ens-eq}. Similarly,
\begin{align}
\label{eq:P-cond}
p(I_{\ell_k} | I_{\ell_{k-1}}) =
 \frac{1 + I_{\ell_k} \nb_{\ell_k}\rb_{\rm ens}(t_k\big| I_{\ell_{k-1}}\nb_{\ell_{k-1}}, t_{k-1})}{2},
\end{align}
where $\rb_{\rm ens}(t_k \big| I_{\ell_{k-1}}\nb_{\ell_{k-1}}, t_{k-1})$ is the qubit state at time $t_k$ obtained from Eq.~\eqref{eq:ens-eq} with the initial condition $\rb =I_{\ell_{k-1}}\nb_{\ell_{k-1}}$ at the time $t_{k-1}$. This initial condition is  due to collapse of the qubit state by projective measurement of $\sigma_{\ell_{k-1}}$ at the time $t_{k-1}$ with result $I_{\ell_{k-1}}$.
Note that while the collapse recipe is applicable to correlators, it is not applicable to the joint probability distribution of the continuous output signals $I_{\ell_k}(t_k)$, for which no formula like Eq.\ \eqref{eq:collapse-recipe} is possible.

{\it Correlator factorization for unital evolution}.---Using  Eqs.~\eqref{eq:collapse-recipe} and \eqref{eq:P-cond}, we can write the $N$-time correlator as
\begin{align}
\label{eq:main-result-deriv}
&K_{\ell_1...\ell_N}(t_1,...\, t_N) = \sum\nolimits_{\{{I}_{\ell}=\pm1\}}^{2^{N-2}} \sum\nolimits_{I_{\ell_{N-1}}=\pm 1}    I_{\ell_{N-1}}
    \nonumber \\
&\hspace{0.4cm}\times \nb_{\ell_N} \rb_{\rm ens} (t_N \big|{I}_{\ell_{N-1}}\nb_{\ell_{N-1}},t_{N-1})
    \nonumber \\
&\hspace{0.4cm} \times [1 + {I}_{\ell_{N-1}} \nb_{\ell_{N-1}} \rb_{\rm ens} (t_{N-1}\big|{I}_{\ell_{N-2}}\nb_{\ell_{N-2}},t_{N-2})]/2
    \nonumber\\
&\hspace{0.4cm} \times p(I_{\ell_{N-2}}\big|I_{\ell_{N-3}})\cdots p(I_{\ell_2}\big|I_{\ell_1})\, p(I_{\ell_1})\prod\nolimits_{k=1}^{N-2}I_{\ell_k},
\end{align}
where we separated the factors for the latest pair of measurements and already summed over the result $I_{\ell_N}=\pm 1$ of the latest measurement. For the summation over $I_{\ell_{N-1}}=\pm 1$, let us use the  symmetry property (\ref{eq:unital-prop}) of unital evolution, which gives $\rb_{\rm ens}(t_N | I_{\ell_{N-1}}\nb_{\ell_{N-1}},t_{N-1}) = I_{\ell_{N-1}}\rb_{\rm ens}(t_N | \nb_{\ell_{N-1}},t_{N-1})$. This cancels the factor $I_{\ell_{N-1}}$ on the first line of Eq.\ (\ref{eq:main-result-deriv}) since $I_{\ell_{N-1}}^2=1$, so summation over $I_{\ell_{N-1}}$ only affects the second term in the third line, which sums to zero. Therefore, the last two measurements bring only the factor $\nb_{\ell_N} \rb_{\rm ens}(t_N | \nb_{\ell_{N-1}},t_{N-1})$, which does not depend on the previous measurement results. Moreover, for unital evolution, this factor is exactly the two-time correlator $K_{\ell_{N-1} \ell_N}(t_{N-1},t_N)$, as easy to see using the collapse recipe. We emphasize that for unital evolution the two-time correlators do not depend on the initial state \cite{Atalaya2017corr}. Thus, for unital evolution we obtain factorization
    \begin{align}
& K_{\ell_1...\ell_N}(t_1,...\, t_N) = K_{\ell_1...\ell_{N-2}}(t_1,...\, t_{N-2})
    \nonumber \\
&\hspace{3cm} \times K_{\ell_{N-1}\ell_N}(t_{N-1},t_N) .
    \label{eq:2-factorization} \end{align}
Continuing this procedure, we see that for even $N$ the $N$-time correlator is a product of two-time correlators,
         \begin{align}
 K_{\ell_1...\ell_N}(t_1,...\, t_N) = \prod\nolimits_{i=1}^{N/2}
K_{\ell_{2i-1}\ell_{2i}}(t_{2i-1},t_{2i}),
    \label{eq:even-N} \end{align}
while for odd $N$ the remaining factor is the average output signal at $t_1$, which depends on the initial state,
   \begin{align}
 K_{\ell_1...\ell_N}(t_1,...\, t_N) = \langle I_{\ell_1} (t_1) \rangle \prod\nolimits_{i=1}^{\frac{N-1}{2}}
K_{\ell_{2i}\ell_{2i+1}}(t_{2i},t_{2i+1}).
    \label{eq:odd-N} \end{align}
The two-time correlators do not depend on the initial state and can be calculated by integrating Eq.\ (\ref{eq:ens-eq}) (with $\rb_{\rm st}=0$ for unital evolution),
\begin{align}
\label{eq:two-time-corr}
K_{\ell_{i}\ell_{k}}(t_i,t_k) =  \nb_{\ell_k}\left[ \exp \left( \int_{t_i}^{t_k}\Lambda_{\rm ens}(t)\, dt \right) \nb_{\ell_{i}}\right],
\end{align}
where the exponential is time-ordered. For a time-independent $\Lambda_{\rm ens}$, the two-time correlator (\ref{eq:two-time-corr}) is obviously a function of the time difference $t_k-t_i$, while not depending on $t_i$.

The multi-time correlator factorization (\ref{eq:even-N}) and (\ref{eq:odd-N}) in the case of unital evolution is the {\it main result} of this paper. Rather surprisingly, the $N$-time correlator does not depend on the qubit evolution between neighboring time moments $t_{N-2i}$ and $t_{N-2i+1}$; in particular, the time duration between them is not important. For even $N$, the correlator also does not depend on the initial qubit state. 

Our factorization result may seem similar to Wick's theorem in Gaussian field theory~\cite{KamenevBook}, indicating a trivial correlation. However, this is not the case, since in our problem only some specific pairwise correlators contribute to Eqs.\ (\ref{eq:even-N}) and (\ref{eq:odd-N}), while others do not. This is a rather peculiar correlation; for example, the evolution between time moments $t_{N-2i}$ and $t_{N-2i+1}$ obviously affects the joint probability distribution but cancels out in the correlator.

Note that the factorization result is somewhat similar to the result of Ref.\ \cite{Diosi2016} for sequential infinitesimally weak measurements with no evolution in between. In contrast, we show that the factorization holds for continuous measurements in the presence of an arbitrary unital evolution (but without phase backaction). If the qubit evolution is not unital, then the $N$-time correlator still can be calculated via Eq.\ (\ref{eq:collapse-recipe}) using the collapse recipe. However, the correlator does not factorize. A recursive relation for the $N$-time correlator in this case is presented in Supplemental Material \cite{Supplement}.

{\it Singular contributions at coinciding times.}---So far we assumed that all time moments $t_k$ in the correlator \eqref{eq:Kn-def} are different.
If a pair of neighboring time moments, $t_k$ and $t_{k+1}$, approach each other and they correspond to the same detector, $\ell_k=\ell_{k+1}$, then we also need to add to the correlator a singular contribution, $\propto \delta (t_{k+1}-t_k)$, due to white noise in this detector -- see Eq.\ (\ref{eq:noise-corr}). The additional contribution will be  $\tau_{\ell_k}\delta(t_{k+1}-t_k)\, K_{N-2}$, where $K_{N-2}$ denotes the remaining $(N-2)$-time correlator with the coinciding pair excluded. Similarly, if there are two pairs of coinciding times (time-separated from each other), which involve the same detectors,  $\ell_k=\ell_{k+1}$,  $\ell_i=\ell_{i+1}$, then there will also be a contribution $\tau_{\ell_k}\tau_{\ell_i}\delta(t_{k+1}-t_k)\, \delta(t_{i+1}-t_i)\, K_{N-4}$, where $K_{N-4}$ is the $(N-4)$-time correlator without coinciding pairs (if $\ell_k \neq \ell_i$, then the formula is the same even if these two pairs coincide in time). These formulas for the singular contributions do not assume unital evolution. Note that there is no singular contributions from
three coinciding times (with the same detector) because the noises $\xi_\ell(t)$ are assumed to be Gaussian.

{\it Comparison with experiment.---}To check our theoretical results for multi-time correlators, we use the data from the experiment described in detail in Ref.~\cite{Shay2016} (two-time correlators have been analyzed in Ref.\ \cite{Atalaya2017corr}). In this experiment, two linear detectors measure simultaneously and continuously the observables $\sigma_z$ and $\sigma_\varphi = \sigma_z\cos\varphi + \sigma_x \sin \varphi$ of a nominally non-evolving qubit, with corresponding normalized output signals $I_z(t)$ and $I_\varphi(t)$. Here $\varphi$ is the angle between the measurement axes in the Bloch $xz$ plane of the measured qubit; in the experiment $\varphi=n\pi/10$ with integer $n$ from 0 to 10 (we neglect the small correction $\Delta \varphi = 0.036$ \cite{Atalaya2017corr}). As shown in Ref.~\cite{Atalaya2017corr}, the decoherence-caused evolution of the measured effective qubit is unital even in the presence of energy relaxation of the physical qubit (because of averaging over sufficiently fast Rabi oscillations of the physical qubit, creating the effective rotating-frame qubit). Since in the experiment a Josephson parametric amplifier operated in phase-sensitive mode has been used, amplifying the informational (optimal) quadrature,  the phase backaction is nominally absent. Therefore, all conditions for our factorization result for multi-time correlators, Eqs.~\eqref{eq:even-N}--\eqref{eq:two-time-corr}, are satisfied in the  experiment.

\begin{figure}[t]
\centering
\begin{tabular}{cc}
\includegraphics[width=0.8\linewidth, trim = 2cm 0.2cm 2cm 0cm,clip=true]{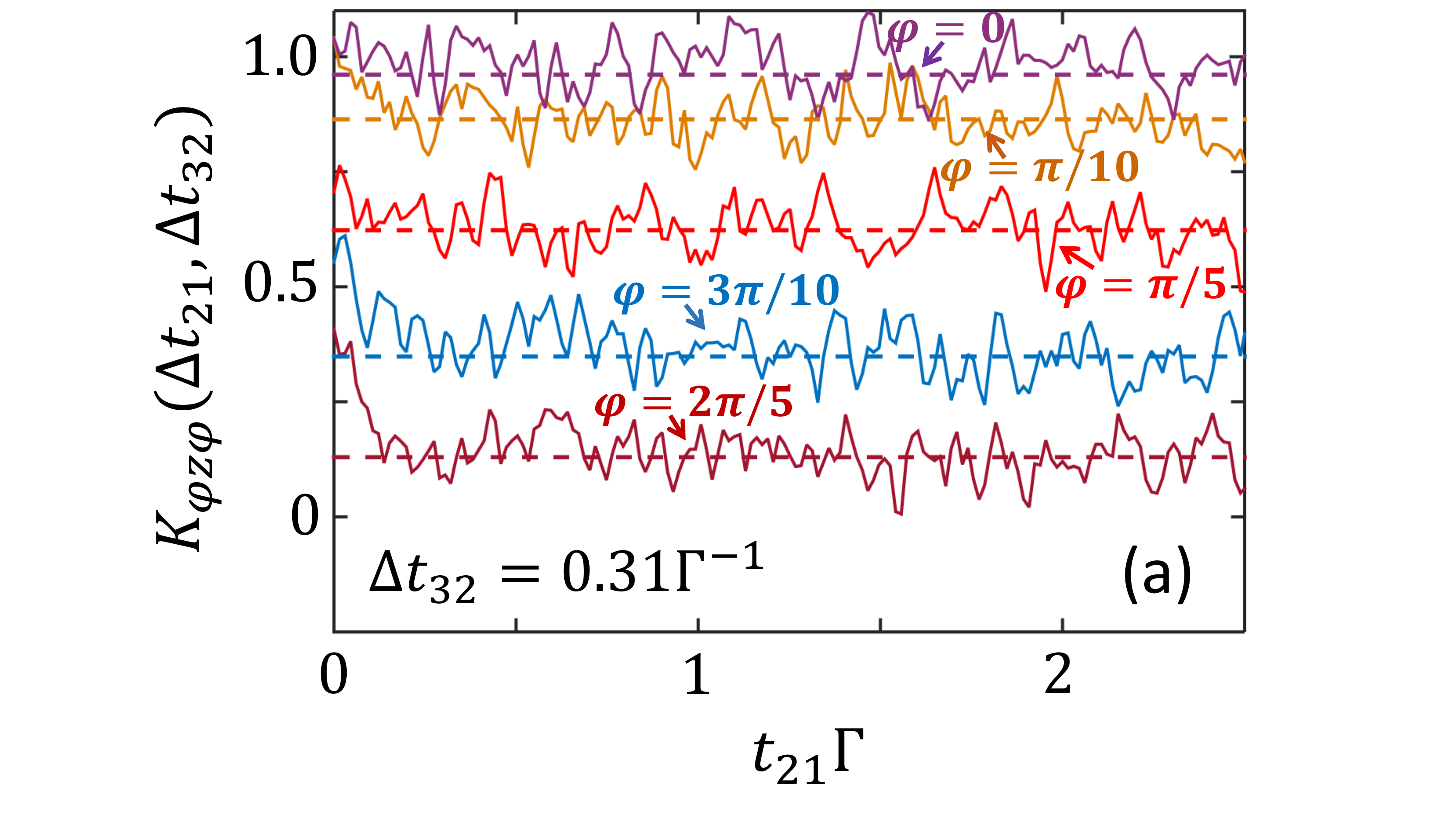} \\
\includegraphics[width=0.8\linewidth, trim = 2cm 0.75cm 2cm 0cm,clip=true]{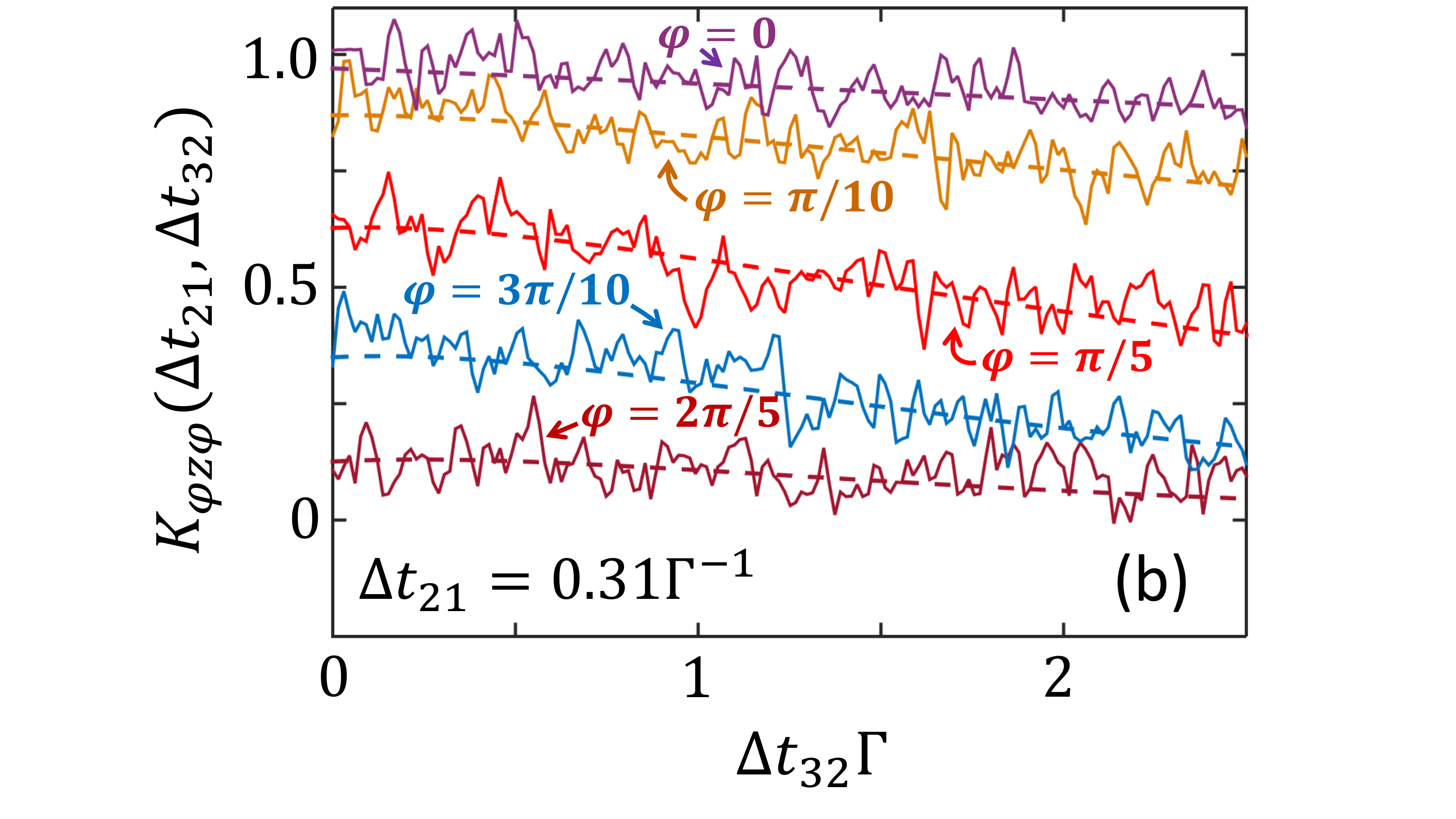}
\end{tabular}
\caption{Comparison between experimental (solid lines) and theoretical (dashed lines) three-time correlators for simultaneous measurement of qubit observables $\sigma_z$ and $\sigma_\varphi$, with $\varphi$ being the angle between the measurement axes on the Bloch sphere. Upper and lower panels show the correlator $K_{\varphi z\varphi}(\Delta t_{21},\Delta t_{32})$  as a function of the time difference $\Delta t_{21}$ and $\Delta t_{32}$, respectively (see text). As predicted by theory, we see practically no dependence on $\Delta t_{21}$, in contrast to a significant dependence on $\Delta t_{32}$. In both measurement channels, $\Gamma = (1.3\,\mu{\rm s})^{-1}$. }
\label{fig:3time-corr}
\vspace{-0.5cm}
\end{figure}

Let us first consider the three-time correlator $K_{\varphi z\varphi}(\Delta t_{21},\Delta t_{32}) \equiv \int_{t_a}^{t_a+T}dt_1  \, \langle I_\varphi(t_3) \, I_z(t_2)\, I_\varphi(t_1)\rangle/T$, where $\Delta t_{21}=t_2-t_1$, $\Delta t_{32}=t_3-t_2$, and additional averaging over time $t_1$ within the interval $[t_a, t_a+T]$ is introduced to reduce fluctuations of the experimental correlators. For brevity of notation, we omit dependence of $K_{\varphi z\varphi}$ on $t_a$ and $T$; in Fig.\  \ref{fig:3time-corr} we choose $t_a=1\,\mu$s and $T=0.2\,\mu$s. As follows from (\ref{eq:odd-N}), $K_{\varphi z\varphi}(\Delta t_{21},\Delta t_{32})=K_{z\varphi}(\Delta t_{32})\int_{t_a}^{t_a+T}dt_1\, \langle I_{\varphi}(t_1)\rangle/T$, where the two-time correlator $K_{z\varphi}(\Delta t_{32})$ and the average signal $\langle I_{\varphi}(t)\rangle= z(t)\cos\varphi + x(t)\sin\varphi$ can be found using the results of Ref.\ \cite{Atalaya2017corr}, using the qubit initial state $\rb(0) = \{\sin(\varphi/2),0,\cos(\varphi/2)\}$ -- see Section B of Supplemental Material \cite{Supplement}. Note that each measurement channel produces  the  measurement-induced ensemble dephasing rate of $\Gamma=(1.3\, \mu{\rm s})^{-1}$ in the corresponding basis. Our theory predicts no dependence of $K_{\varphi z\varphi}$ on $\Delta t_{21}$. In agreement with this prediction, the experimental correlators $K_{\varphi z\varphi}$ shown by solid lines in Fig.~\ref{fig:3time-corr}(a) for several values of $\varphi$, do not exhibit a significant dependence on $\Delta t_{21}$. Some deviations from the theory (dashed lines) at  $\Delta t_{21} < 0.5\, \Gamma^{-1}$ are probably due to slowly fluctuating offsets of the output signals. In contrast, the theory predicts a significant dependence of $K_{\varphi z\varphi}$ on $\Delta t_{32}$, which also agrees with experimental correlators shown in Fig.~\ref{fig:3time-corr}(b). Note that experimental three-time correlators are much noisier than the two-time correlators discussed in Ref.~\cite{Atalaya2017corr}, so the ensemble averaging over 200,000  experimental trajectories still produces significant fluctuations in Fig.~\ref{fig:3time-corr} (see \cite{Supplement} for details of experimental signal processing).

\begin{figure}[tb]
\centering
\begin{tabular}{cc}
\includegraphics[width=0.85\linewidth, trim = 0cm 1cm 0cm 0cm,clip=true]{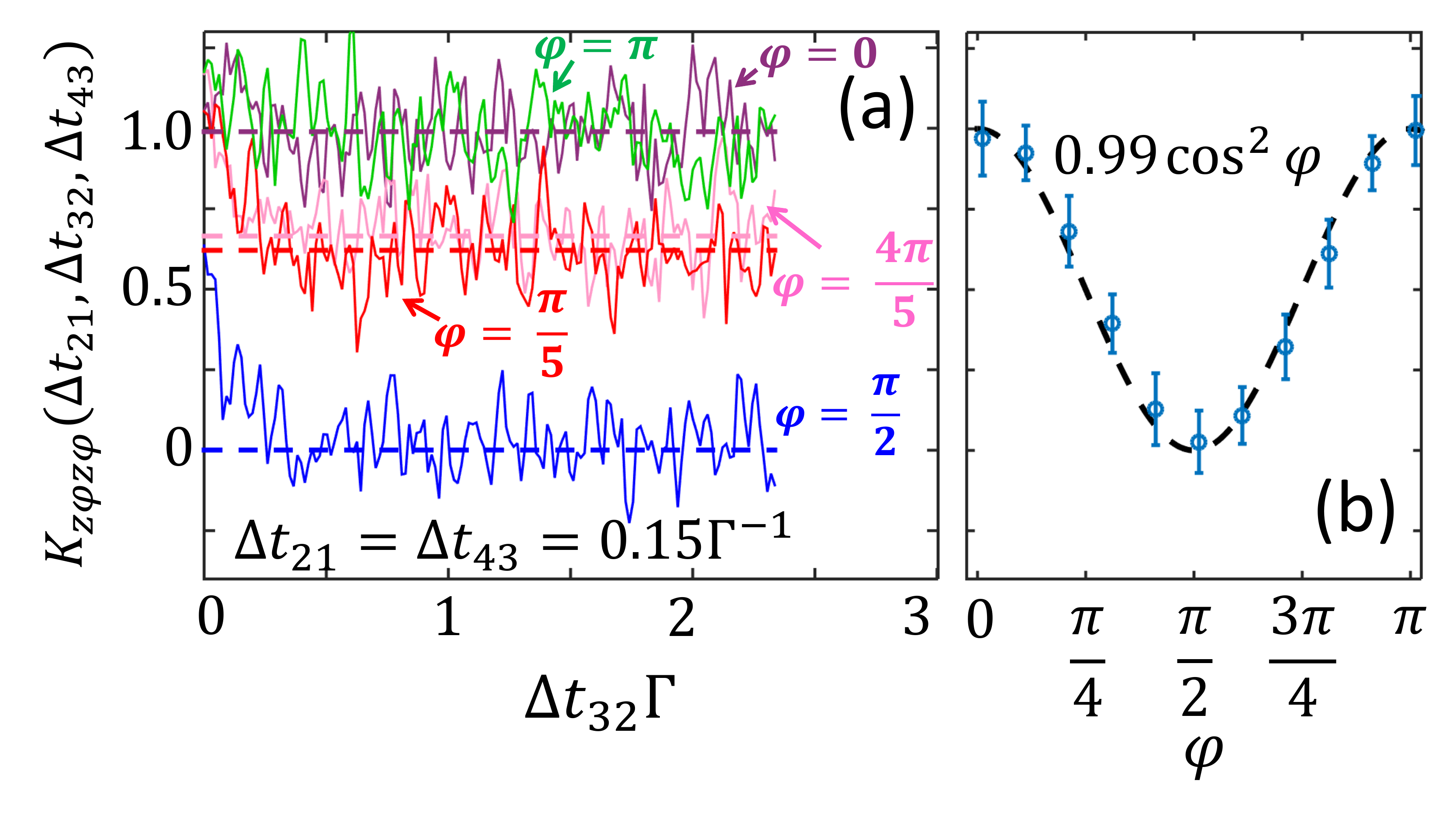}\\
\includegraphics[width=0.8\linewidth, trim = 1.0cm 0.5cm 2.0cm 0cm,clip=true]{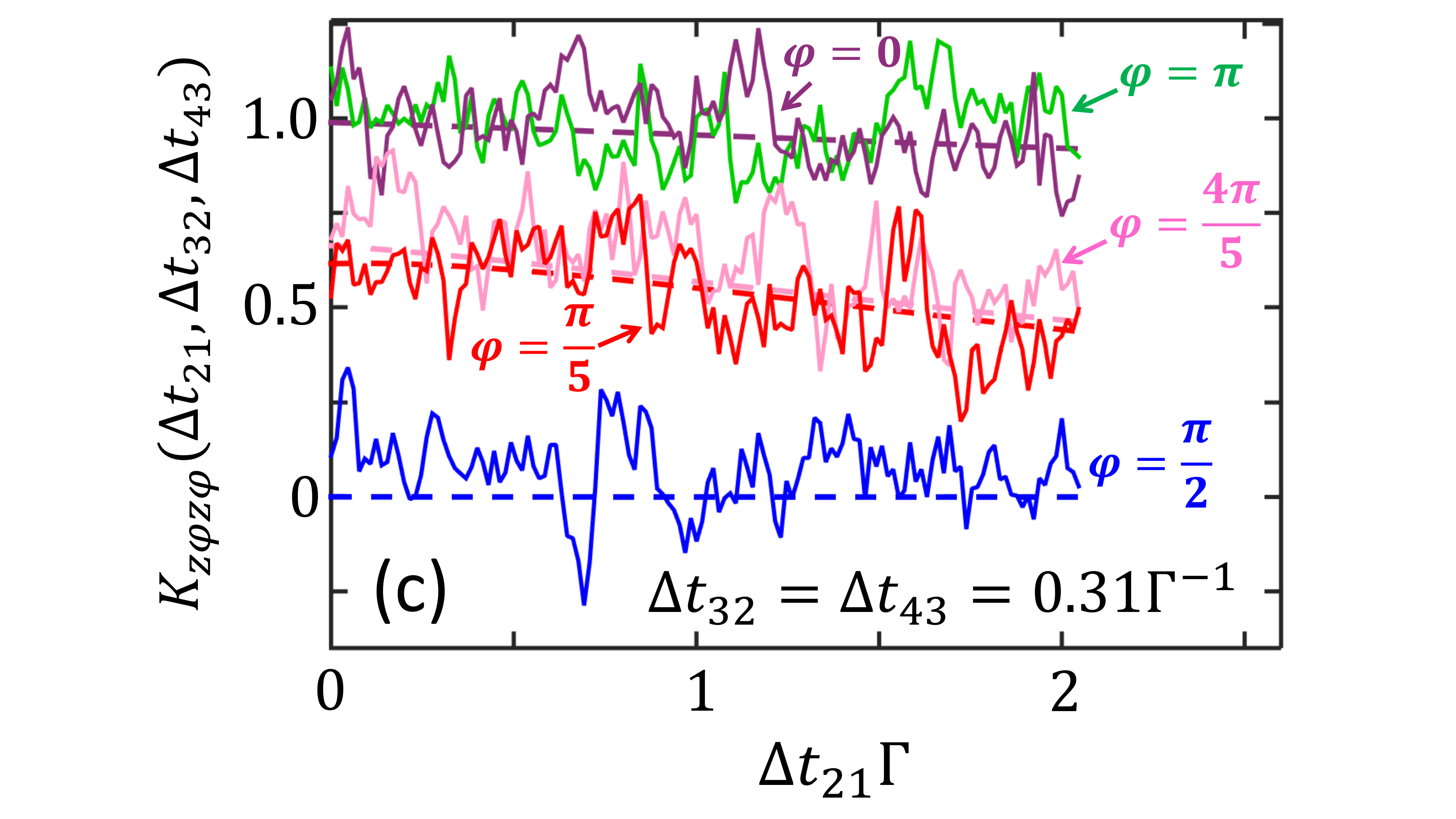}
\end{tabular}
\caption{Comparison between experimental and theoretical four-time correlators. Panels (a) and (c) depict the dependence of the correlator $K_{z\varphi z\varphi}(\Delta t_{21},\Delta t_{32},\Delta t_{43})$ on  $\Delta t_{32}$ and $\Delta t_{21}$, respectively (solid lines for experimental results, dashed lines for theory). Theory predicts no dependence on $\Delta t_{32}$.  Panel (b) depicts the average (circles) and standard deviation (error bars) of the experimental correlators shown in panel (a), averaged over $\Delta t_{32}$ within the range [$0.5/ \Gamma, 2.3/\Gamma$], for several values of $\varphi$. The dashed line in (b) is the theoretical result $0.99 \,\cos^2\varphi$.
}
\label{fig:4time-corr}
\vspace{-0.5cm}
\end{figure}

Next, let us consider the similar four-time correlator $K_{z\varphi z\varphi}(\Delta t_{21},\Delta t_{32},\Delta t_{43})\equiv \int_{t_a}^{t_a+T}dt_1\,\langle I_\varphi(t_4) \, I_z(t_3) \times  I_\varphi(t_2) \, I_z(t_1)\rangle/T$ with $\Delta t_{ij}=  t_{i}- t_{j}$. As follows from Eq.\ (\ref{eq:even-N}), $K_{z\varphi z\varphi}(\Delta t_{21},\Delta t_{32},\Delta t_{43})= K_{z\varphi}(\Delta t_{21})\, K_{z\varphi}(\Delta t_{43})$, predicting that $K_{z\varphi z\varphi}$ should not depend on the time difference $\Delta t_{32}$ (it should also not depend on $t_a$ and $T$).
Figure~\ref{fig:4time-corr}(a) shows the dependence of the experimental correlators $K_{z\varphi z\varphi}$ on $\Delta t_{32}$ for several values of $\varphi$, with $\Delta t_{21}=\Delta t_{43}=0.15\, \Gamma^{-1}$, $t_a= 1\,\mu$s, and $T=0.5 \, \mu$s. Indeed, we see that experimental $K_{z\varphi z\varphi}$ fluctuate around the theoretical constant values (horizontal dashed lines), except for $\Delta t_{32}< 0.5\, \Gamma^{-1}$, where the solid lines deviate up from the theory [probably because of the same reason as in Fig.\ \ref{fig:3time-corr}(a)]. Figure~\ref{fig:4time-corr}(b) shows the same values of $K_{z\varphi z\varphi}$ averaged over $\Delta t_{32}$ (excluding $\Delta t_{32}<0.5\, \Gamma^{-1}$) as a function of $\varphi$. We see a good agreement with the theoretical result (shown by dashed line), $K_{z\varphi}^2(\Delta t)= A\cos^2\varphi$, where $A\approx 0.99$ for $\Delta t=0.15\, \Gamma^{-1}$. Figure~\ref{fig:4time-corr}(c) depicts the dependence of the experimental correlator on $\Delta t_{21}$ (with fixed $\Delta t_{32}$ and $\Delta t_{43}$); we again see a good agreement with the theory. Similar results have been obtained for the dependence on $\Delta t_{43}$ (not shown).
Note that the four-time correlators in Fig.~\ref{fig:4time-corr}(a,c) are even noisier than the three-time correlators in Fig.~\ref{fig:3time-corr}; in general, higher-order correlators are increasingly noisier because of multiplication of noise terms.

{\it Conclusion.---}We have analyzed multi-time correlators for the output signals of linear detectors, continuously measuring arbitrary observables of a qubit at the same time. We have shown that an $N$-time correlator can be expressed as a simple product of two-time correlators for even $N$ [Eq.~\eqref{eq:even-N}], while for odd $N$ there is also a factor equal to the average signal at the earliest time [Eq.~\eqref{eq:odd-N}]. This result requires the absence of the phase backaction from continuous measurements and also requires a unital ensemble-averaged evolution. Experimental results for three-time and four-time correlators show good agreement with the theory. Our results can be used in parameter estimation and noise detection protocols in qubit systems, and also for development and analysis of quantum error correction codes operating with continuous measurements; for this purpose a generalization of our theory to multiqubit systems may be needed.

{\it Acknowledgements.---}We thank Andrew Jordan for useful discussions. The work was supported by ARO grant No. W911NF-15-1-0496. L.S.M acknowledges
support from the National Science Foundation Graduate Fellowship grant No. 1106400.

\appendix
\setlength{\parindent}{10pt}
\begin{center}
\center{ {\bf \center Section A: Proof of the ``collapse recipe'' } }
\end{center}
\vspace{0.3cm}
\setcounter{equation}{1}
In this section we prove the ``collapse recipe'', which states that in the absence of phase backaction, the multi-time correlator
\begin{align}\tag{S\theequation}
\label{eq:sup-Kn-def}
K_{\ell_1 \ell_2...\ell_N}(t_1,t_2,...t_N)\equiv \big\langle I_{\ell_N}(t_N). . .\, I_{\ell_2}(t_2)\, I_{\ell_1}(t_1)\big\rangle
\end{align}\stepcounter{equation}%
can be calculated by replacing the actual continuous measurement with projective measurement of operators $\sigma_{\ell_k}$ at time moments $t_k$ ($k=1, 2, ... \, N$, $t_1<t_2< ... <t_N$), while the qubit evolution at $t\neq t_k$ is replaced with the ensemble-averaged evolution. In the proof we show that the value of the correlator $K$ obtained in this way coincides with the value obtained from the quantum Bayesian formalism, in which the qubit evolution is described by the stochastic equation (in It\^o interpretation)
    \begin{align}\tag{S\theequation}
\label{eq:sup-Bayesian-eq-v2}
\dot\rb = \Lambda_{\rm ens}(\rb - \rb_{\rm st}) + \sum_{\ell=1}^{N_{\rm d}} \frac{\nb_\ell - (\nb_\ell\rb)\, \rb}{\sqrt{\tau_\ell}}\, \xi_\ell(t),
    \end{align}\stepcounter{equation}%
where $\rb$ is the vector of Bloch-sphere components, $\rho=\openone /2+\rb \bsigma /2$, the output signal of the $\ell$th detector continuously measuring the qubit operator $\sigma_\ell = \bn_\ell \bsigma$  is
    \begin{align}\tag{S\theequation}
\label{eq:sup-I-ell}
I_\ell(t) = \bn_\ell \rb(t) + \sqrt{\tau_\ell}\, \xi_\ell(t),
    \end{align}\stepcounter{equation}%
$\tau_\ell$ is the corresponding measurement (collapse) time (the quantum efficiency $\eta_\ell$ is not important for correlators), and $\xi_\ell(t)$ are the uncorrelated white noises with $\langle \xi_\ell(t)\, \xi_{\ell'}(t')\rangle = \delta_{\ell\ell'}\delta(t-t')$. The Markovian ensemble-averaged evolution of the qubit state is given by Eq.\ (\ref{eq:sup-Bayesian-eq-v2}) without the noise term,
    \begin{align}\tag{S\theequation}
\label{eq:sup-ens-eq}
      \dot{\rb}_{\rm ens} = \Lambda_{\rm ens}(\rb_{\rm ens} - \rb_{\rm st}).
    \end{align}\stepcounter{equation}%
The evolution is assumed to start with some initial state $\rb_{\rm in}$  at time $t_{\rm in}\leq t_1$.
All parameters of the measurement and evolution ($\Lambda_{\rm ens}$, $\rb_{\rm st}$, $\nb_\ell$, $\tau_\ell$, $\eta_{\ell}$)  can be time-dependent. 

\indent We will first prove the collapse recipe in a simple way and then will prove it in another, more formal way.

\subsubsection*{1. Simple proof}

The simple proof of the collapse recipe closely follows the proof for two-time correlators in Refs.\ \cite{S-Korotkov2001sp} and \cite{S-Atalaya2017corr}. Understanding of this proof is easier after understanding of proofs in Refs.\ \cite{S-Korotkov2001sp} and \cite{S-Atalaya2017corr}.

The proof uses linearity of quantum mechanics. In particular, from the linearity, the correlator (\ref{eq:sup-Kn-def}) can depend on the initial state $\rb_{\rm in}$ only linearly, $K=\vb \rb_{\rm in}+C$, where the vector $\vb$ and the number $C$ can depend on all parameters for the correlator, but do not depend on $\rb_{\rm in}$. The linearity is better seen by introducing 4-vectors for unnormalized density matrices, $\tilde\rb=(u,x,y,z)$ for $\rho= (u\openone +x\sigma_x+y\sigma_y+z\sigma_z)/2$; then $K=\tilde \vb \tilde\rb_{\rm in}$ with some 4-vector $\tilde\vb$, which does not depend on $\tilde\rb_{\rm in}$. Note that quantum evolution is linear for 4-vectors $\tilde\rb$, but is not necessarily linear for 3-vectors $\rb$. The evolution, which is linear for 3-vectors $\rb$, is called unital.

The correlator (\ref{eq:sup-Kn-def}) is the average over the ensemble of quantum trajectories, starting with initial state $\rb_{\rm in}$ at time $t_{\rm in}$. Let us discretize time into small but still non-zero timesteps $\Delta t$, so that the noises $\xi_\ell (t)$ are not infinitely large ($|\xi_\ell| \sim 1/\sqrt{\Delta t}$). Since the values of the output signals $I_\ell (t)$ at $t\neq t_k$ do not affect the correlator $K$, we can pretend that during these timesteps the signals $I_\ell (t)$ are not available to any observer, and therefore the qubit evolution is equivalent \cite{S-Korotkov2001} to ensemble-averaged evolution given by Eq.\ (\ref{eq:sup-ens-eq}). Thus, we need to take into account the full Bayesian evolution (\ref{eq:sup-Bayesian-eq-v2}) only during timesteps $t_k$. Moreover,
since at time $t_k$ only the output from $\ell_k$th detector affects the correlator, in Eq.\ (\ref{eq:sup-Bayesian-eq-v2}) we can neglect all the terms in the sum except for $\ell=\ell_k$. Integrating Eq.\ (\ref{eq:sup-Bayesian-eq-v2}) over the timestep $\Delta t$ around $t_k$, we obtain the ``Bayesian kick''
    \begin{equation}\tag{S\theequation}
    \Delta \rb_k \equiv \Delta \rb (t_k)= \frac{\nb_{\ell_k} - ( \nb_{\ell_k} \rb_k)\, \rb_k}{\sqrt{\tau_{\ell_k}}}\, \xi_{\ell_k}(t_k) \, \Delta t ,
    \label{eq:sup-kick}\end{equation}\stepcounter{equation}%
where $\rb_k\equiv \rb (t_k)$ and $|\xi_{\ell_k}|\sim 1/\sqrt{\Delta t}$. With $\Delta t\to 0$, this information-induced kick becomes infinitesimally small, so its effect on further evolution is infinitesimally small. However,  its contribution to the correlator (\ref{eq:sup-Kn-def}) is significant, since the signal $I_{\ell_k}(t_k)$ in the correlator contains the term $\sqrt{\tau_{\ell_k}}\, \xi_{\ell_k}(t_k)$ [see Eq.\ (\ref{eq:sup-I-ell})], so the effect of the Bayesian kick (\ref{eq:sup-kick}) is proportional to the product
    \begin{equation}\tag{S\theequation}
    \sqrt{\tau_{\ell_k}}\, \xi_{\ell_k}(t_k) \Delta \rb_k = [\nb_{\ell_k} - ( \nb_{\ell_k} \rb_k)\, \rb_k] \,  \xi_{\ell_k}^2(t_k) \, \Delta t,
    \label{eq:sup-product}\end{equation}\stepcounter{equation}%
which is non-zero since on average $ \xi_{\ell_k}^2(t_k) \, \Delta t=1$.

Let us prove that we can apply the collapse recipe to the measurement at time $t_k$ in the correlator (\ref{eq:sup-Kn-def}). This means that the value of the correlator would not change if we replace the actual signal $I_{\ell_k}(t_k)$ in Eq.\ (\ref{eq:sup-Kn-def}) by $I_{\ell_k}(t_k)=\pm 1$ with probabilities
    \be\tag{S\theequation}
    p_k^{\pm} = \frac{1\pm \nb_{\ell_k} \rb_k}{2},
    \label{eq:sup-prob}\ee\stepcounter{equation}%
and correspondingly greatly increase the Bayesian kick by starting the further evolution with the state $\rb(t_k+0)=\pm \nb_{\ell_k}$ (i.e., the corresponding eigenstate of the measured operator $\sigma_{\ell_k}$). In the proof we assume fixed (though arbitrary) values for all previous measurements $I_{\ell_{k'}}(t_{k'<k})$, so that $\rb_k$ is fixed. Then the $N$-time correlator (\ref{eq:sup-Kn-def}) reduces to a product of $I_{\ell_1}(t_1)\, I_{\ell_2}(t_2)\, ...\, I_{\ell_{k-1}}(t_{k-1})$ and the remaining $(N+1-k)$-time correlator. Therefore, this correlator depends linearly on $\rb_k$ (better to say, on 4-vector $\tilde \rb_k$ -- see discussion above).

Let us separate the correlator (\ref{eq:sup-Kn-def}) [with fixed  $I_{\ell_{k'}}(t_{k'<k})$] into two terms, $K=K_{k}^{(1)}+K_{k}^{(2)}$, which correspond to the two terms in Eq.\ (\ref{eq:sup-I-ell}) at time $t_k$, i.e., $K_{k}^{(1)}=\langle I_{\ell_N}(t_N) ...I_{\ell_{k+1}}(t_{k+1})\rangle \, \nb_{\ell_k} \rb_k \, I_{\ell_{k-1}} ...I_{\ell_1}(t_1)$ and $K_{k}^{(2)}=\langle I_{\ell_N}(t_N) ...I_{\ell_{k+1}}(t_{k+1})  \sqrt{\tau_{\ell_k}}\, \xi_{\ell_k}(t_k) \rangle \, I_{\ell_{k-1}} ...I_{\ell_1}(t_1)$. Because of the quantum linearity, the value of $K_{k}^{(1)}$ will not change if we replace $\nb_{\ell_k} \rb_k$ with $I_{\ell_k}=+1$ and start the further evolution with the unnormalized density matrix $\rho_1^+ (t_k+0)= (\nb_{\ell_k} \rb_k)\, \rho (t_k)= (\nb_{\ell_k} \rb_k)\, \openone /2 + (\nb_{\ell_k} \rb_k) (\rb_k \bsigma /2)$. Note that we need to multiply all elements of $\rho$ by $\nb_{\ell_k} \rb_k$; this is why the normalization changes, ${\rm Tr}(\rho_1^+)=\nb_{\ell_k} \rb_k$. This is necessary because for non-zero stationary state $\rb_{\rm st}$, the evolution (\ref{eq:sup-ens-eq}) of the Bloch-sphere components is non-linear, even though the evolution of the density matrix $\rho$ is linear.

The same linearity-based idea for $K_{k}^{(2)}$ needs to take into account the Bayesian kick (\ref{eq:sup-kick}). It is easy to see that  $K_{k}^{(2)}$ will not change if we replace $\sqrt{\tau_{\ell_k}}\, \xi_{\ell_k}(t_k)$ with $I_{\ell_k}=+1$ and start the further evolution with $\rho_2^+ (t_k+0)=  [\nb_{\ell_k} - ( \nb_{\ell_k} \rb_k)\, \rb_k] \, \bsigma /2$ -- see Eq.\ (\ref{eq:sup-product}). Note the zero trace of $\rho_2^+$; this is because the Bayesian kick does not change the trace.

Adding the contributions from $K_{k}^{(1)}$ and $K_{k}^{(2)}$ and using the linearity, we see that the correlator $K$ will not change if we replace $I_{\ell_k}$ with  $I_{\ell_k}=+1$ and start the further evolution with $\rho^+(t_k+0)=\rho_1^+(t_k+0)+\rho_2^+(t_k+0)= (\nb_{\ell_k} \rb_k) \, \openone /2 + \nb_{\ell_k} \bsigma /2$. Using the linearity again, we see that $K$ will also not change if we replace  $I_{\ell_k}$ with $I_{\ell_k}=-1$ and start the further evolution with $\rho^-(t_k+0)=-\rho^+(t_k+0)= -(\nb_{\ell_k} \rb_k)\, \openone /2 - \nb_{\ell_k} \bsigma /2$.
The value of $K$ will also not change if we use one of these two replacements probabilistically. Note that $\rho^\pm (t_k+0)$ differ from the normalized eigenstates $\openone/2\pm \nb_{\ell_k} \bsigma /2$ of the measured operator $\sigma_{\ell_k}$ only by $(\pm \nb_{\ell_k} \rb_k -1) \, \openone /2$. If we choose the replacements $I_{\ell_k}=\pm 1$ with probabilities given by Eq.\ (\ref{eq:sup-prob}), then the effect of this difference will be cancelled on average since $\sum_{\pm} \pm (\pm \nb_{\ell_k} \rb_k -1)(1\pm \nb_{\ell_k} \rb_k)=0$. Therefore, the value of $K$ does not change if we start the further evolution with the states $\openone/2\pm \nb_{\ell_k} \bsigma /2$, as if after the standard projective measurement of $\sigma_{\ell_k}$.

Thus, we have proven that we can apply the collapse recipe to the measurement at time $t_k$, assuming fixed measurement results for the previous measurements. Since the values of the previous measurement results are arbitrary, the assumption of fixed results is not needed. Finally, since the collapse recipe can be applied separately to measurement at any time moment $t_k$ in the correlator (\ref{eq:sup-Kn-def}), it can be applied to all of them. This completes the proof of the collapse recipe for multi-time correlators (\ref{eq:sup-Kn-def}).

\vspace{0.5cm}
Note that instead of using the collapse recipe and working with normalized states, we can also calculate the correlator using the described above procedure based on unnormalized states. In this case at each moment $t_k$, we replace $I_{\ell_k}(t_k)$ with $I_{\ell_k}=+1$ and start the further evolution with $\rho^+(t_k+0)=(\nb_{\ell_k} \rb_k)\, u_k \, \openone /2 + \nb_{\ell_k} \bsigma /2$, where $u_k={\rm Tr}[\rho (t_k)]$ accounts for possibly unnormalized state $\rho$ before $t_k$. Since this procedure can be applied for all $N$ moments $t_k$ and then the product of all $I_{\ell_k}$ is 1, the value of the correlator is simply the norm of the state after the last time moment $t_N$. Therefore this new ``one-path recipe'' for calculating the $N$-time correlator (\ref{eq:sup-Kn-def}) is the following. Start with the initial (normalized) state $\tilde{\bf r}_{\rm in}$ at the initial time $t_{\rm in}$ and propagate it using the ensemble-averaged evolution (\ref{eq:sup-ens-eq}) (which does not change the norm), also adding the ``state jumps'' (which change the norm) at time moments $t_k$ as $\rho(t_k+0)=(\nb_{\ell_k} \rb_k)\, u_k \,\openone /2 + \nb_{\ell_k} \bsigma /2$. Then the norm of the resulting state $\rho (t_N+0)$ is the value of the correlator.

This one-path recipe can be easily generalized to arbitrary measurement operators in an arbitrary system. For a continuous measurement of an arbitrary Hermitian observable $A$, the quantum Bayesian evolution due to informational backaction (in the absence of a unitary backaction) is (the derivation is simple, the result is the same as in the Quantum Trajectory theory \cite{Wiseman-book})
    \begin{align}\tag{S\theequation}
\hspace{0.2cm}    \dot{\rho} = \frac{A\rho A -(A^2\rho+\rho A^2)/2}{2\eta S} +\frac{A\rho +\rho A - 2\rho {\rm Tr} (A\rho)}{\sqrt{2S}}\, \xi(t),
    \label{eq:Bayes-general}\end{align}\stepcounter{equation}%
where $\xi (t)$ is the normalized white noise, $\langle \xi (t)\,\xi(t')\rangle =\delta (t-t')$, extracted from the normalized detector signal, $I(t)={\rm Tr}(A\rho) + \sqrt{S/2}\, \xi(t)$, $S$ is the single-sided spectral density of the detector signal noise, and $\eta$ is the quantum efficiency (so that the fraction $\eta$ of the noise $S$ is ``quantum-limited''). Since the first term in Eq.\ (\ref{eq:Bayes-general}) is obviously the ensemble-averaged (Lindblad) evolution, the evolution due to measurement of several (generally non-commuting) observables $A_\ell$ in the presence of additional unitary evolution and decoherence (but still without unitary backaction from measurement) is
\begin{align}\tag{S\theequation}
\hspace{-0.0cm} \dot{\rho} =  {\cal L} [\rho] + \sum\nolimits_\ell \frac{A_\ell \rho +\rho A_\ell - 2\rho {\rm Tr} (A_\ell\rho)}{\sqrt{2S_\ell}}\, \xi_\ell(t),
\label{eq:Bayes-general-2}
\end{align}\stepcounter{equation}%
where $I_\ell (t)={\rm Tr}(A_\ell \rho)+\sqrt{S_\ell /2} \, \xi_\ell (t)$, $\langle \xi_\ell (t)\, \xi_{\ell '}(t')\rangle =\delta_{\ell \ell'}\, \delta(t-t')$, and ${\cal L}[\rho ]$ is the ensemble-averaged Lindbdad evolution, with the contribution from measurement ${\cal L}_{\rm m}[\rho ]= \sum_\ell [A_\ell \rho A_\ell -(A_\ell^2\rho+\rho A_\ell^2)/2]/(2\eta_\ell S_\ell)$. Following the same idea as described above, for each time moment $t_k$ in the correlator (\ref{eq:sup-Kn-def}), we can remove $I_{\ell_k}(t_k)$ from the correlator $K$,  separating it into two parts, $K=K_k^{(1)}+K_k^{(2)}$, so that for $K_k^{(1)}$ the (unnormalized) state jumps from $\rho_k=\rho(t_k)$ to $\rho_1(t_k+0)={\rm Tr}(A_{\ell_k} \rho_k) \, \rho_k$, while for $K_k^{(2)}$ the ``Bayesian kick'' changes $\rho_k$ into $\rho_2(t_k+0)= (A_{\ell_k}\rho_k +\rho_k A_{\ell_k} )/2 - {\rm Tr}(A_{\ell_k}\rho_k)\, \rho_k$ -- see Eq.\ (\ref{eq:Bayes-general-2}). Therefore, we can remove $I_{\ell_k}(t_k)$ from the correlator (\ref{eq:sup-Kn-def}), replacing it with the state jump
    \be\tag{S\theequation}
\label{eq:rho-k}
\rho(t_k+0) = (A_{\ell_k}\rho_k +\rho_k A_{\ell_k} )/2.
    \ee\stepcounter{equation}%

Thus, the one-path recipe for the $N$-time correlator (\ref{eq:sup-Kn-def}) in the general case is
\begin{align}
&\hspace{0.3cm} K_{\ell_1 \ell_2...\ell_N}(t_1,t_2,...t_N) =    {\rm Tr}\big[ {\cal M}_{t_N}\,  {\cal E}(t_N |t_{N-1})\, {\cal M}_{t_{N-1}} . . .
    \nonumber \\
& \hspace{2.8cm}   {\cal M}_{t_2} \, {\cal E}(t_2 |t_1) \, {\cal M}_{t_{1}} \, {\cal E}(t_1 |t_{\rm in})\, \rho_{\rm in} \big],
\label{eq:one-path-gen}\tag{S\theequation} 
\end{align}\stepcounter{equation}%
where ${\cal E}(t|t')$ is the trace-preserving ensemble-averaged evolution (operation) from time $t'$ to $t$ due to Lindblad term $\dot{\rho} ={\cal L}[\rho]$, while  ${\cal M}_{t_k}\rho = (A_{\ell_k}\rho +\rho A_{\ell_k})/2$ is the trace-changing operation, related to measurement (without unitary backaction) of the operator $A_{\ell_k}$ at time $t_k$.
If a unitary backaction of the form $\sum_\ell -i[B_\ell,\rho] \, \xi_\ell (t)/\sqrt{2S_\ell}$ is added into Eq.\ (\ref{eq:Bayes-general-2}) ($B_\ell$ are Hermitian), with the contribution to the ensemble-averaged evolution absorbed by ${\cal L}[p]$, then the additional Bayesian kick produces  an extra term in Eq.\ (\ref{eq:rho-k}): ${\cal M}_{t_k}\rho = (A_{\ell_k}\rho +\rho A_{\ell_k})/2-i[B_{\ell_k},\rho]/2$.
The one-path recipe is similar to the result of a recent paper \cite{Tilloy2017} by Tilloy. Note the similarity of Eq.\ (\ref{eq:one-path-gen}) to  the quantum regression formula.

The one-path recipe (\ref{eq:one-path-gen}) based on unnormalized states can be reduced to the physically transparent collapse recipe (based on physical states) only when $B_\ell=0$ and $A_\ell^2$ are positive numbers, i.e., scaled unity operators.  (In the general case, it is still possible to generalize the collapse recipe to work with normalized, but unphysical states; however, then the physical meaning becomes obscure.)  In particular, the collapse recipe is fully applicable for continuous measurement of multi-qubit Pauli operators in an arbitrary system of qubits, because then $A_\ell^2=\openone$. One can see this by noticing that Eq.~\eqref{eq:rho-k} in this case can be written as $\rho(t_k+0) = \sum_{\pm} \pm  {\rm Tr} [\rho_k\Pi_{\ell_k}^\pm ] \left(\displaystyle{\frac{\Pi_{\ell_k}^\pm  \rho_k\Pi_{\ell_k}^\pm}{{\rm Tr}[ \rho_k\Pi_{\ell_k}^\pm] }} \right)$, where $\Pi_{\ell_k}^\pm$ is the projection operator corresponding to the eigenvalue $\pm 1$ of $A_{\ell_k}$. This form corresponds to the result $\pm 1$ of the projective measurement of $A_{\ell_k}$, with probability ${\rm Tr} [\rho_k\Pi_{\ell_k}^\pm ]$ and with the density matrix inside the parenthesis being the normalized state after the projective multi-qubit collapse.

Completing the brief digression into the general case, we remind that the main purpose of this section is the proof of the collapse recipe for the case of a single qubit, considered in this paper.

\subsubsection*{2. Alternative proof}

Now let us prove the collapse recipe for the single-qubit case in a different, more formal way. In this derivation we will also obtain the correlator factorization result for unital evolution, Eq.\ (11) of the main text.

In addition to the correlator $K$ given by Eq.\ (\ref{eq:sup-Kn-def}), let us introduce the vector-valued correlator
\begin{align}\tag{S\theequation}
\label{eq:Kn-vector}
\hspace{0.3cm} \Kb_{\ell_1...\ell_N}(t_1,...\, t_N) \equiv \big\langle \rb(t_N) \, I_{\ell_{N-1}}(t_{N-1}) \cdots I_{\ell_1}(t_1)\big\rangle .
\end{align}\stepcounter{equation}%
Note that in this notation for $\Kb$, the last subscript $\ell_N$ is not needed, but we keep it to remind us that $\Kb$ is an average product of  $N$ terms. We will usually assume $t_1<t_2<...<t_N$ (as for the correlator $K$), but at some point in the derivation we will need the time moment $t_N$ to cross $t_{N-1}$. The correlator $K$ can be easily obtained from $\Kb$ as
\begin{align}\tag{S\theequation}
\label{eq:Kn-v2}
K_{\ell_1...\ell_N} (t_1,...\, t_N) = \nb_{\ell_N} \Kb_{\ell_1...\ell_N}(t_1,...\, t_N),
\end{align}\stepcounter{equation}%
since the noise contribution $\sqrt{\tau_{\ell_N}} \, \xi_{\ell_{N}}(t_N)$ to the output signal $I_{\ell_N}(t_N)$ [see Eq.\ (\ref{eq:sup-I-ell})] is not correlated with past qubit states.

Let us separate $\Kb$ into two terms, $\Kb=\Kb^{(1)}+\Kb^{(2)}$, which correspond to the two terms in Eq.\ (\ref{eq:sup-I-ell}) for the signal $I_{\ell_{N-1}}(t_{N-1})$,
    \begin{align}
\label{eq:Kn-1}
&\hspace{0.1cm}\Kb_{\ell_1...\ell_N}^{(1)}(t_1,...\, t_N)  \equiv \big\langle \rb(t_N)\big[\nb_{\ell_{N-1}}\rb(t_{N-1}) \big]
    \nonumber\\
&\hspace{3.2cm} \times I_{\ell_{N-2}}(t_{N-2}) \cdots I_{\ell_1}(t_1)\big\rangle,
   \tag{S\theequation}     \\
\label{eq:Kn-2}
&\hspace{0.1cm}\Kb_{\ell_1...\ell_N}^{(2)}(t_1,...\, t_N)  \equiv
 \big\langle \rb(t_N)  \big[\sqrt{\tau_{\ell_{N-1}}}\, \xi_{\ell_{N-1}}(t_{N-1}) \big]
    \nonumber\\
&\hspace{3.2cm} \times I_{\ell_{N-2}}(t_{N-2})\cdots I_{\ell_1}(t_1)\big\rangle.{\stepcounter{equation} \tag{S\theequation}}
\end{align}\stepcounter{equation}%
The derivative of $\Kb^{(1)}$ over the last time moment $t_N$ can be
obtained from Eq.~\eqref{eq:sup-Bayesian-eq-v2},
\begin{align}
\label{eq:Kn-1-eqm}
&\hspace{0.3cm}\partial_{t_N} \Kb_{\ell_1...\ell_N}^{(1)}(t_1,...t_N) =
\Lambda_{\rm ens} (t_N) \, \Big( \Kb_{\ell_1...\ell_N}^{(1)}(t_1,...t_N)
    \nonumber \\
&\hspace{2.5cm} - \rb_{\rm st} (t_N) \, K_{\ell_1...\ell_{N-1}}(t_1,...t_{N-1}) \Big),
\tag{S\theequation}\end{align}\stepcounter{equation}%
where we included possible dependence of $\Lambda_{\rm ens}$ and $\rb_{\rm st}$ on time. The initial condition at $t_N=t_{N-1}+0$ is
\begin{align}
\label{eq:Kb1-in-cond}
&\hspace{0.3cm}\Kb_{\ell_1...\ell_N}^{(1)}(t_1,...\, t_{N-1},t_{N-1}+0) = \big\langle  \rb (t_{N-1})
     \nonumber \\
 & \hspace{1.0cm}
\times \big( \nb_{\ell_{N-1}}\rb(t_{N-1})\big) \, I_{\ell_{N-2}}(t_{N-2})\cdots I_{\ell_1}(t_1)\big\rangle.\tag{S\theequation}
\end{align}\stepcounter{equation}%
The time derivative of $\Kb^{(2)}$ over $t_N$ can also be
obtained from Eq.~\eqref{eq:sup-Bayesian-eq-v2}, which gives
\begin{align}
\label{eq:Kn-2-eqm}
&\hspace{0.2cm}\partial_{t_N} \Kb_{\ell_1...\ell_N}^{(2)}(t_1,...\, t_N) = \Lambda_{\rm ens}(t_N) \, \Kb_{\ell_1...\ell_N}^{(2)}(t_1,...\, t_N)
    \nonumber  \\
&\hspace{0.8cm}+\Big\langle\big[\nb_{\ell_{N-1}} - \rb(t_{N-1}) \big( \nb_{\ell_{N-1}} \rb(t_{N-1}) \big)\big]
   \nonumber \\
&\hspace{1.4cm} \times I_{\ell_{N-2}}(t_{N-2}) \cdots I_{\ell_1}(t_1)\Big\rangle \, \delta (t_N-t_{N-1}).\tag{S\theequation}
\end{align}\stepcounter{equation}%
Note that $\Kb^{(2)}=0$ for $t_N<t_{N-1}$ because of causality, so the second term in Eq.\ (\ref{eq:Kn-2-eqm}) sets the initial condition at $t_N=t_{N-1}+0$, caused by the Bayesian kick. Also note that at $t_N>t_{N-1}$, the evolution of $\Kb^{(2)}$ is linear (due to $\Lambda_{\rm ens}$); it does not have the inhomogeneous term containing $\rb_{\rm st}$ as for the evolution of $\Kb^{(1)}$ in Eq.\ (\ref{eq:Kn-1-eqm}).

Solving Eqs.\ \eqref{eq:Kn-1-eqm}--\eqref{eq:Kn-2-eqm}, we find $\Kb^{(1)}$ and $\Kb^{(2)}$ for $t_N>t_{N-1}$, starting with the value (\ref{eq:Kb1-in-cond}) of $\Kb^{(1)}$ at $t_N=t_{N-1}+0$,
\begin{align}
\label{eq:Kn-1-sol}
&\hspace{0.2cm} \Kb_{\ell_1...\ell_N}^{(1)}(t_1,...\,t_N) =
 {\mathcal P}(t_N | t_{N-1})\,\Kb_{\ell_1...\ell_N}^{(1)}(t_1,...\,t_{N-1}+0)
    \nonumber \\
& \hspace{1.0cm} +  {\boldsymbol {\mathcal P}}_{\rm st}(t_N | t_{N-1})\, K_{\ell_1...\ell_{N-1}}(t_1,...\, t_{N-1}) ,
   \tag{S\theequation} \\
&\hspace{0.2cm} \Kb_{\ell_1...\ell_N}^{(2)}(t_1,...\,t_N) =  - {\mathcal P}(t_N |t_{N-1}) \Kb_{\ell_1...\ell_N}^{(1)}(t_1,...\,t_{N-1}+0)
    \nonumber \\
&\hspace{1.0cm} +  {\mathcal P} (t_N |t_{N-1})\, \nb_{\ell_{N-1}} K_{\ell_1...\ell_{N-2}}(t_1,...\,t_{N-2}),{\tag{S\theequation}\stepcounter{equation}}
\label{eq:Kn-2-sol}
\end{align}\stepcounter{equation}%
where ${\mathcal P}(t|t')$ is the 3$\times$3 propagator matrix for the homogeneous part of the ensemble-averaged evolution (\ref{eq:sup-ens-eq}), so that $\partial_t {\mathcal P}(t|t') = \Lambda_{\rm ens}(t)\, {\mathcal P}(t|t')$ for $t>t'$ and ${\mathcal P} (t|t)=\openone$, while ${\boldsymbol {\mathcal P}}_{\rm st}(t|t')$ is the contribution from the inhomogeneous part,
 \be\tag{S\theequation}
{\boldsymbol {\mathcal P}}_{\rm st}(t|t')= - \int_{t'}^{t}  {\mathcal P}(t|t'')\, \Lambda_{\rm ens}(t'') \, \rb_{\rm st}(t'') \, dt'' .
\label{eq:sup-P-st}\ee\stepcounter{equation}

From Eqs.~\eqref{eq:Kn-1-sol}--\eqref{eq:sup-P-st}  and~\eqref{eq:Kn-v2}--\eqref{eq:Kn-2} we find the recursive formula, which relates the $N$-time correlator $K_{\ell_1...\ell_N}(t_1,...\,t_N)$ with $(N-1)$-time correlator and $(N-2)$-time correlator (for $N>2$)
\begin{align}
\label{eq:Kn-recur-rel}
&\hspace{0.2cm} K_{\ell_1...\ell_N}(t_1,...\, t_{N}) = \nb_{\ell_N} \,
{\mathcal P}(t_N |t_{N-1})\, \nb_{\ell_{N-1}}
    \nonumber \\
& \hspace{1.2cm}\times  K_{\ell_1...\ell_{N-2}}(t_1,...\,t_{N-2})\, \nonumber \\
& \hspace{1.0cm} +  \nb_{\ell_N}  {\boldsymbol {\mathcal P}}_{\rm st}(t_N | t_{N-1}) \, K_{\ell_1...\ell_{N-1}}(t_1,...\,t_{N-1})  .\tag{S\theequation}
\end{align}\stepcounter{equation}%
Note that for $N=2$, the only difference in the derivation is that the product  $I_{\ell_{N-2}}(t_{N-2}) \cdots I_{\ell_1}(t_1)$ in Eq.\ (\ref{eq:Kn-2-eqm}) should be replaced with 1. As a consequence, the $(N-2)$-time correlator in Eqs.\ (\ref{eq:Kn-2-sol}) and (\ref{eq:Kn-recur-rel}) should be replaced with 1. Therefore, Eq.\ (\ref{eq:Kn-recur-rel}) is also valid for $N=2$ if we define the $0$-time correlator as being equal to 1.

Thus, the recursive relation (\ref{eq:Kn-recur-rel}) is sufficient to derive explicit formulas for $N$-time correlators, if we complement it with the correlator for $N=1$, which is simple,
\begin{align}\tag{S\theequation}
\label{eq:Kell1}
K_{\ell_1}(t_1) = \nb_{\ell_1} \rb(t_1) .
\end{align}\stepcounter{equation}
Now let us show that the $N$-time correlators obtained via Eqs.\ (\ref{eq:Kn-recur-rel}) and (\ref{eq:Kell1}) coincide with the correlators obtained using the collapse recipe. Since for $N=1$ the collapse recipe obviously gives Eq.\ (\ref{eq:Kell1}), we only need to prove that the recursive relation (\ref{eq:Kn-recur-rel}) also follows from the collapse recipe (with the correlator for $N=0$ defined as 1). Note that applicability of the collapse recipe to the two-time correlator was proven in Ref.\ \cite{S-Atalaya2017corr}.

Let us rewrite Eq.\ (7) of the main text (following from the collapse recipe, as indicated by the superscript below) in the form
\begin{align}
\label{eq:deriv-collapse-1}
&\hspace{0.3cm} K_{\ell_1...\ell_N}^{\rm coll}(t_1,...\, t_N) = \sum\nolimits_{\{{I}_{\ell_k}=\pm1\}}^{2^N} {I}_{\ell_N}
 \nonumber \\
& \hspace{0.2cm} \times \frac{1 + {I}_{\ell_N}\nb_{\ell_N} \rb_{\rm ens}\big(t_N\big|{I}_{\ell_{N-1}}\nb_{\ell_{N-1}},t_{N-1}\big)}{2}
    \nonumber \\
&\hspace{0.2cm} \times \prod\nolimits_{k=2}^{N-1} \big[ {I}_{\ell_{k}}p\big({I}_{\ell_k},t_{k}\big|{I}_{\ell_{k-1}},t_{k-1}\big)\big] \times {I}_{\ell_1}p({I}_{\ell_1},t_1),\tag{S\theequation}
 \end{align}\stepcounter{equation}%
where
  \begin{align}
\label{eq:ens-eq-sol}
&\rb_{\rm ens}\big(t_N\big|{I}_{\ell_{N-1}}\nb_{\ell_{N-1}},t_{N-1}\big) =I_{\ell_{N-1}} {\mathcal P}(t_N|t_{N-1})\, \nb_{\ell_{N-1}}
    \nonumber \\
&\hspace{4.4cm} + {\boldsymbol {\mathcal P}}_{\rm st}(t_N|t_{N-1}),\tag{S\theequation}
 \end{align}\stepcounter{equation}%
is the solution of the ensemble-averaged evolution \eqref{eq:sup-ens-eq} with the initial condition ${I}_{\ell_{N-1}}\nb_{\ell_{N-1}}$ at time $t_{N-1}$. Note that the last line of Eq.\ (\ref{eq:deriv-collapse-1}) summed over all combinations of $I_{\ell_k}=\pm 1$ except summation over $I_{\ell_N}$, is the $(N-1)$-time correlator $K_{\ell_1...\ell_{N-1}}^{\rm coll}(t_1,...\, t_{N-1})$.

The term 1 in the second line of Eq.\ (\ref{eq:deriv-collapse-1}) can be removed because of summation over $I_{\ell_N}=\pm 1$. After removing 1,  we see that $I_{\ell_N}$ in the first and second lines cancel each other since $I_{\ell_N}^2=1$. Therefore, Eq.\ (\ref{eq:deriv-collapse-1}) can be rewritten as
\begin{align}
\label{eq:deriv-collapse-add}
&\hspace{0.3cm} K_{\ell_1...\ell_N}^{\rm coll}(t_1,...\, t_N) = \sum\nolimits_{\{{I}_{\ell_k}=\pm1\}}^{2^{N-1}} \nb_{\ell_N}
 \nonumber \\
& \hspace{0.3cm} \times \big[   I_{\ell_{N-1}}
{\mathcal P}(t_N|t_{N-1})\, \nb_{\ell_{N-1}}
+ {\boldsymbol {\mathcal P}}_{\rm st}(t_N|t_{N-1}) \big]
    \nonumber \\
&\hspace{0.3cm} \times \prod\nolimits_{k=2}^{N-1} \big[ {I}_{\ell_{k}}p\big({I}_{\ell_k},t_{k}\big|{I}_{\ell_{k-1}},t_{k-1}\big)\big] \times {I}_{\ell_1}p({I}_{\ell_1},t_1),\tag{S\theequation}
 \end{align}\stepcounter{equation}%
where there is already no summation over the last output $I_{\ell_N}$, and we used Eq.\ (\ref{eq:ens-eq-sol}) for $\rb_{\rm ens}$. Let us separate $K^{\rm coll}$ into two terms, corresponding to contributions from the two terms in the second line of Eq.\ \eqref{eq:deriv-collapse-add}. The second term (containing $ {\boldsymbol {\mathcal P}}_{\rm st}$) is $\nb_{\ell_N} {\boldsymbol {\mathcal P}}_{\rm st}(t_N|t_{N-1}) K_{\ell_1...\ell_{N-1}}^{\rm coll}(t_1,...\, t_{N-1})$, thus coinciding with the third line of Eq.\ (\ref{eq:Kn-recur-rel}). In the remaining first term, let us substitute the product $\prod_{k=2}^{N-1}$ with product $\prod_{k=2}^{N-2}$ multiplied by the corresponding factor for $k=N-1$, and then use relations $I_{\ell_{N-1}}^2=1$ and $\sum_{I_{\ell_{N-1}}=\pm 1} p(I_{\ell_{N-1}},t_{N-1}|I_{\ell_{N-2}},t_{N-2})=1$. This gives us  $\nb_{\ell_N} {\mathcal P}(t_N|t_{N-1})\, \nb_{\ell_{N-1}} K_{\ell_1...\ell_{N-2}}^{\rm coll}(t_1,...\, t_{N-2})$, which is the first term in Eq.\ (\ref{eq:Kn-recur-rel}). Thus, we have obtained
the same recursive relation  (\ref{eq:Kn-recur-rel}) for $K^{\rm coll}$. Therefore, we have proven that the collapse recipe gives the same result for $N$-time correlators as the calculation based on the stochastic evolution equation (\ref{eq:sup-Bayesian-eq-v2}).

Note that the recursive relation (\ref{eq:Kn-recur-rel}) can be used directly to derive the main result of the paper: factorization of the $N$-time correlator in the case of unital evolution, Eq.\ (11) of the main text. Since $\rb_{\rm st}=0$ for unital evolution, from Eq.\ (\ref{eq:sup-P-st}) we obtain ${\boldsymbol {\mathcal P}}_{\rm st}=0$, so the recursive formula (\ref{eq:Kn-recur-rel}) relates the $N$-time correlator only with the $(N-2)$-time correlator. It is easy to see that the coefficient
$\nb_{\ell_N}  {\mathcal P}(t_N |t_{N-1})\, \nb_{\ell_{N-1}}$ is the two-time correlator $K_{\ell_{N-1}\ell_N}(t_{N-1},t_N)$, as also follows from Eq.\ (\ref{eq:Kn-recur-rel}) for $N=2$, since $K=1$ for $N=0$. Thus, for unital evolution, the $N$-time correlator is a product of two-time correlator for the two latest time moments and the remaining $(N-2)$-time correlator. This gives  Eq.\ (11) of the main text.

\vspace{1.5cm}



\begin{center}
{\bf \center Section B: Experimental multi-time correlators}
\end{center}
\vspace{0.3cm}

Our theoretical results for the correlators have been checked against experimental data from the experiment, in which a physical qubit (transmon), embedded into a 3D Al cavity, is subject to relatively fast Rabi oscillations with frequency $\Omega_{\rm R}/2\pi = 40$ MHz. The physical qubit is dispersively coupled to the two lowest cavity modes; each of them is off-resonantly driven with two sideband tones at frequencies $\omega_{{\rm r},i}\pm \Omega_{\rm rf}$ (where $\Omega_{\rm rf}\approx \Omega_{\rm R}$), with a relative phase $\delta_i$. Here $i=z,\varphi$ labels the cavity mode that performs continuous measurement of the observable $\sigma_i$ of the effective (rotating frame) qubit, and $\omega_{{\rm r},i}$ is the frequency of the corresponding cavity mode. Details of the measurement technique are discussed in Ref.\ \cite{S-Shay2016} (see also Ref.\ \cite{S-Atalaya2017corr}).

The measured effective qubit is defined in the frame, which rotates with frequency $\Omega_{\rm rf}$ with respect to the laboratory frame of the physical qubit.
The measurement axes on the Bloch sphere of the effective qubit are determined by the relative phases $\delta_i$ of the sideband tones, with position of the effective $z$ axis defined arbitrarily within the $xz$ plane of the physical qubit rotations. We choose one of the measurements to be exactly the $\sigma_z$ measurement; the other measurement direction is shifted by an angle $\varphi$, thus corresponding to the observable $\sigma_\varphi \equiv \sigma_z \cos\varphi + \sigma_x\sin\varphi$.
In the experiment $\varphi=n\pi/10$ with integer $n=0,\, 1, \, 2, ...\,10$. The effective qubit is initialized at $t=0$ in the middle between the measurement axes, i.e., at the states  $\rb_0^{\pm}=\pm\{\sin(\varphi/2),0,\cos(\varphi/2)\}$. Approximately 200,000 readout trajectories are recorded for each angle $\varphi$, with approximately 100,000 trajectories for each initial state (we use only  trajectories, selected by heralding the ground state at the start of a run and checking that transmon is still within the two-level subspace after the run \cite{S-Atalaya2017corr}).

The ensemble-averaged evolution for the effective qubit is  \cite{S-Atalaya2017corr}
    \begin{align}
& \hspace{0.3cm}  \dot x= - \Gamma_z x - \Gamma_{\varphi}  \cos\varphi \left( x\cos\varphi - z\sin\varphi \right) +\tilde{\Omega}_{\rm}z - \gamma x,
    \label{eq:eq-x} \tag{S\theequation}\\
& \hspace{0.3cm}  \dot y= - (\Gamma_z +\Gamma_\varphi + T_2^{-1})\, y,
    \label{eq:eq-y} {\tag{S\theequation} \stepcounter{equation}}\\
& \hspace{0.3cm}  \dot z= \Gamma_{\varphi}  \sin\varphi \left(x\cos\varphi - z\sin\varphi\right)  -\tilde{\Omega}_{\rm}x - \gamma z,
\label{eq:eq-z} {\tag{S\theequation} \stepcounter{equation}}
\end{align}\stepcounter{equation}%
where $\Gamma_z$ and $\Gamma_\varphi$ are the measurement-induced dephasing rates in the corresponding bases of the two measurement channels, $\tilde \Omega_{\rm R} = \Omega_{\rm R} -\Omega_{\rm rf}$ is a small residual Rabi oscillation frequency, $\gamma = (T_1^{-1} + T_2^{-1})/2$, and  $T_{1}$ and $T_2$ are the intrinsic energy relaxation and dephasing times for the physical qubit. In the experiment $\Gamma_z\approx \Gamma_\varphi\approx (1.3\,\mu{\rm s})^{-1}$ (denoted $\Gamma$ in the main text), $T_1=60\,\mu$s, $T_2=30\,\mu$s, and $\tilde \Omega_{\rm R}/2\pi \simeq 12\,$kHz.

Experimental three-time correlators are calculated using the experimental (unnormalized and slightly shifted) output signals  $\tilde I_z(t)$ and $\tilde I_\varphi (t)$ from the two measurement channels as
\begin{align}
\label{eq:K3-exp}
&  \hspace{0.3cm} K^{\pm}_{\varphi z \varphi}(\Delta t_{21},\Delta t_{32}) = \int_{t_a}^{t_a+T} \Big\langle \frac{\tilde I_\varphi(t+\Delta t_{21}+\Delta t_{32}) - \tilde I_\varphi^{\rm off}}{\Delta \tilde I_\varphi /2 }
\nonumber\\
& \hspace{1.3cm}\times \frac{\tilde I_z(t+\Delta t_{21}) - \tilde I_z^{\rm off}}{\Delta \tilde I_z/2} \,\, \frac{ \tilde I_\varphi (t) - \tilde I_\varphi^{\rm off}\big)}{\Delta \tilde I_\varphi /2}\Big\rangle \, \frac{dt}{T} , \tag{S\theequation}
\end{align}\stepcounter{equation}%
where the time integration over duration $T = 0.2\,\mu$s is needed to reduce correlator fluctuations, the small constant offsets  $\tilde I_{z,\varphi}^{\rm off}$ are less than 0.2 in magnitude (see \cite{S-Atalaya2017corr} for details), and experimental responses are
$\Delta \tilde I_z=4.2$ (in Ref.~\cite{S-Atalaya2017corr} we used 4.0) and $\Delta \tilde I_\varphi=4.4$. To avoid initial transients in the data, we use $t_a=1\,\mu$s. The superscipts in the correlators $K^\pm_{\varphi z\varphi}$ correspond to the initial states $\rb_0^{\pm}$, the ensemble
averaging is over the corresponding $\sim100$,000 trajectories. Since theoretically $K^-_{\varphi z\varphi}=-K^+_{\varphi z\varphi}$, in Fig.~1 of the main text we plot the difference,
\begin{align}
\label{eq:K3-exp-def}
& K_{z\varphi z}(\Delta t_{21},\Delta t_{32})= \big[  K^{+}_{z\varphi z}(\Delta t_{21},\Delta t_{32})
    \nonumber \\
& \hspace{3.2cm} - K^{-}_{z\varphi z}(\Delta t_{21},\Delta t_{32}) \big] /2. \tag{S\theequation}
\end{align}\stepcounter{equation}%

The experimental four-time correlators plotted in Fig.~2 of the main text are calculated  as
\begin{align}
\label{eq:K4-exp}
&\hspace{0.2cm} K_{z\varphi z\varphi}(\Delta t_{21},\Delta t_{32},\Delta t_{43}) = \int_{t_a}^{t_a+T} \Big\langle\frac{\tilde I_\varphi(t+\Delta t_{41}) - \tilde I_\varphi^{\rm off}}{\Delta \tilde I_\varphi /2}
  \nonumber\\
&\hspace{1.5cm} \times \frac{\tilde I_z(t+\Delta t_{31}) - \tilde I_z^{\rm off}}{\Delta \tilde I_z /2 } \,\, \frac{\tilde I_\varphi (t+\Delta t_{21}) - \tilde I_\varphi^{\rm off}}{\Delta \tilde I_\varphi /2}
    \nonumber \\
&\hspace{1.5cm}\times \frac{\tilde I_z(t) - \tilde I_z^{\rm off}}{\Delta \tilde I_z/2}\Big\rangle\, \frac{dt}{T}, \tag{S\theequation}
\end{align}\stepcounter{equation}%
where $\Delta t_{31} = \Delta t_{32} + \Delta t_{21}$, $\Delta t_{41} = \Delta t_{43}+ \Delta t_{31} $, the averaging is now over all trajectories (starting from both $\rb_0^{+}$ and $\rb_0^{-}$), and we use $T=500\, \mu$s and $t_a=1\,\mu$s. We need a larger averaging window $T$ since the four-time correlators are noisier than the three-time correlators.

The theoretical lines in Figs.\ 1 and 2 of the main text are calculated using the two-time correlator $K_{z\varphi}(\tau)=K_{z\varphi}(t_1, t_1+\tau)$ derived in Ref.\ \cite{S-Atalaya2017corr} (see below); for the three-time correlator we also need the average  $\langle I_\varphi(t)\rangle = z(t)\cos\varphi + x(t)\sin\varphi$, which is calculated using Eqs.\ \eqref{eq:eq-x}--\eqref{eq:eq-z} with the initial condition $\rb(0)=\rb_0^\pm$. The two-time correlator $K_{z\varphi}(\tau)$ is calculated analytically as \cite{S-Atalaya2017corr}
\begin{align}
&\hspace{0.4cm}K_{z\varphi}(\tau) = \frac{(\Gamma_z + \Gamma_\varphi) \cos\varphi  + 2\tilde\Omega_{\rm R}\sin\varphi }{2(\Gamma_+ - \Gamma_-)}  \left(e^{-\Gamma_-\tau} -
e^{-\Gamma_+\tau}\right)
    \nonumber \\
&\hspace{1.5cm} +\frac{\cos \varphi}{2}\left( e^{-\Gamma_-\tau} + e^{-\Gamma_+\tau}\right), \tag{S\theequation}
\end{align}\stepcounter{equation}%
 where
\begin{align}
&\hspace{0.4cm}\Gamma_{\pm} =\frac{\Gamma_z +\Gamma_\varphi \pm \big[ \Gamma_z^2 + \Gamma_\varphi^2 +2\Gamma_z\Gamma_\varphi\cos(2\varphi) -4 \tilde\Omega_{\rm R}^2\big]^{1/2} }{2}.  \tag{S\theequation}
\end{align}
\end{thebibliography}%

\end{document}